\documentclass[iop]{emulateapj}

\usepackage{graphicx}
\usepackage{multirow}
\usepackage{amsmath}
\usepackage{xcolor}

\shorttitle{Carbon Chains vs Methanol}
\shortauthors{Graninger et al.}

\renewcommand{\thefootnote}{\fnsymbol{footnote}}
\begin{document}

\title{Carbon Chains and Methanol toward Embedded Protostars\footnote{Based on observations carried out with the IRAM 30m Telescope. IRAM is supported by INSU/CNRS (France), MPG (Germany) and IGN (Spain).}}

\author{Dawn M. Graninger}
\affil{Harvard-Smithsonian Center for Astrophysics, Cambridge, MA, 02138}
\email{dgraninger@cfa.harvard.edu}

\author{Olivia H. Wilkins\altaffilmark{1}}
\affil{Harvard-Smithsonian Center for Astrophysics, Cambridge, MA, 02138}

\and

\author{Karin I. \"{O}berg}
\affil{Harvard-Smithsonian Center for Astrophysics, Cambridge, MA, 02138}

\altaffiltext{1}{Departments of Chemistry and Mathematics, Dickinson College, Carlisle, PA, 17013}

\begin{abstract}
Large interstellar organic molecules are potential precursors of prebiotic molecules. Their formation pathways and chemical relationships with one another and simpler molecules are therefore of great interest. In this paper, we address the relationship between two classes of large organic molecules, carbon chains and saturated complex organic molecules (COMs), at the early stages of star formation through observations of C$_4$H and CH$_3$OH.   We surveyed these molecules with the IRAM 30m telescope toward 16 deeply embedded low-mass protostars selected from the Spitzer $c2d$ ice survey. We find that CH$_3$OH and C$_4$H are positively correlated indicating that these two classes of molecules can coexist during the embedded protostellar stage. The C$_4$H/CH$_3$OH gas abundance ratio tentatively correlates with the CH$_4$/CH$_3$OH ice abundance ratio in the same lines of sight. This relationship supports a scenario where carbon chain formation in protostellar envelopes begins with CH$_4$ ice desorption.

\end{abstract}


\section{Introduction}
Large organic molecules have been widely observed across different stages of star-formation \citep{Herbst09, Sakai13}. Their chemistry is of great interest because large interstellar molecules may serve as precursors to prebiotic chemistry on nascent planets. Large organic molecules also have a great potential as molecular probes. In general, molecular abundance patterns contain information about the past and current environment where they reside. This relationship can be exploited to constrain interstellar environments using molecular line observations, if the chemistry of the target molecule is well understood. Larger molecules present numerous rotational lines at millimeter and centimeter wavelengths and are therefore especially appropriate from an excitation point of view. They suffer, however, from an often poorly constrained formation chemistry.

In this paper we explore how the relationship  of two kinds of larger organic molecules constrain their chemistry. Interstellar organics are  classified as either saturated organic molecules or unsaturated carbon chains. When saturated organic molecules reach sizes of 6 atoms or greater, they are known as complex organic molecules (COMs). These two classes of molecules have been generally supposed to form and exist in very different interstellar and circumstellar environments \citep{Herbst09}.

Carbon chain molecules were first observed in the cold, dark cloud TMC-1 in the form of cyanopolyynes \citep{Little77, Kroto78, Broten78}. In these environments, carbon chains form through efficient low-temperature ion-molecule reactions in the gas phase \citep{Herbst89, Ohishi98}. In 2008, Sakai et al. detected the first evidence of carbon chains in an unexpected region - the warm, inner regions of a low-mass protostar. Detections toward a second source \citep{Sakai09b} suggested that carbon chains are common toward some classes of protostars. The term ``warm carbon chain chemistry" (WCCC) was coined to describe this type of source. Based on low-temperature modeling, ion-molecule chemistry and chemical inheritance are not sufficient to produce the high abundances of carbon chains observed in WCCC sources. \citet{Sakai08} proposed that additional carbon chains can be formed in protostellar envelopes when  methane (CH$_4$), a common interstellar ice, sublimates at 25 K \citep{Oberg08}. In the gas phase, CH$_4$ reacts with C$^+$ to efficiently form carbon chains. This theory has been validated by chemical models \citep{Aikawa12, Hassel11}. 

COMs were initially observed in hot cores present toward high-mass protostars \citep{Blake87}. During the past two decades it has become clear, however, that COMs are abundant in many other circumstellar and interstellar environments as well \cite[see e.g.][]{Cazaux03,Arce08}. COMs have even been detected toward cold cloud cores, the traditional site of formation for carbon chains, \citep{Oberg10a, Bacmann12,Cernicharo12}. Based on experiments and models, COMs form efficiently on grain surfaces through energetic processing of simple ices and are released into the gas phase via thermal or non-thermal processes \citep{Garrod06, Garrod08,Oberg09d}. Recent theoretical work has shown that COMs may also be formed in the gas-phase following the desorption of methanol (CH$_3$OH) \citep{Balucani15}. 

Both models and observations thus reveal that both carbon chains and complex organics can form at a range of temperatures and may therefore coexist in some sources. Based on a small source sample (2 WCCC sources and 3 hot corinos), \citet{Sakai08, Sakai09b} found that carbon chains and COMs are not detected in large abundances in the same low-mass protostellar sources, however. They suggest that this anti-correlation stems from differences in star formation time scales, which results in different compositions of the ice grains in the protostellar envelope, for WCCC vs. hot corino sources. If the starless core phase is short, the bulk of the carbonaceous ice form from accretion of C atoms. The carbon atoms are hydrogenated to form CH$_4$ and the release of CH$_4$ during protostellar formation promotes warm carbon chain chemistry. If the starless core is long-lived, most of the carbon atoms in the gas phase have had time to react and form CO, resulting in a CO dominated ice. The formation of CH$_3$OH is through CO hydrogenation and CH$_3$OH is the starting point of COM chemistry on grains \citep{Sakai13}. 

In this study, we explore the relationship between the COM precursor CH$_3$OH and prototypical carbon chain C$_4$H in a sample of embedded protostars, selected from the Spitzer $c2d$ (cores to disk) ice survey. We further explore the relationship between C$_4$H and CH$_3$OH, and ice abundances in the same lines of sight. In \S2, we describe our source sample and selection criteria. The IRAM 30m observations are described in  \S3. The results and discussion are in \S4 and \S5 and the conclusions can be found in \S6. 

\section{Sample Selection}

Our sources were selected from the Spitzer $c2d$ ice sample \citep{Evans03, Boogert08}. From the 51 sources in the Spitzer $c2d$ ice sample presented in \citet{Boogert08}, we initially constrained ourselves to the northern low-mass sources as they are easily observable with the IRAM 30m, leaving us with 26 sources. The sources are characterized into different classes based on their spectral energy distributions, and in particular their IR spectral indices, $\alpha_{\rm IR}$, defined as the slope between 2 and 24 $\mu$m. We chose sources with $\alpha_{\rm IR}>0.3$, which defines class 0/I sources \citep{Wilking01}. These sources are often, but not always, associated with young, embedded YSOs; this cut left us with 19 sources. We then removed B1-b, IRAS 03301+3111, and EC 92 from our sample. B1-b was removed as previous observations of this source exist, IRAS 03301+3111 was removed as the upper limits on the ice abundances did not place any strong constraints on the data, and EC 92 was removed because it was not resolvable from SVS 4--5 within the beam of the IRAM 30m.

Table \ref{Table1} lists the source coordinates, bolometric luminosities, envelope masses and the IR SED indices, together with the H$_2$O, CH$_4$, and CH$_3$OH ice abundances of our final sample. Our sources span $\alpha_{\rm IR}$ from 0.34 -- 2.70 and are situated in the Perseus, Taurus, Serpens, L1014, and CB244 clouds. The envelope masses are between 0.1 -- 17.7  M$_\odot$, and bolometric luminosities are between 0.32 -- 38 L$_\odot$. Previous observations of WCCC sources fall within our range of bolometric luminosities \citep{Andre00,Chen97}. The H$_2$O ice column is between $0.4-39\times10^{18}$~cm$^{-2}$ and the $N_{\rm CH_3OH}/N_{\rm H_2O}$ and $N_{\rm CH_4}/N_{\rm H_2O}$ are 2--25\% and 1.6--11\%, respectively. Six of our sources (B1-a, B5 IRS 1, L1489 IRS, IRAS 04108+2803, SVS 4-5, and IRAS 03235+3004) were part of a study by \citet{Oberg14a} on COMs, including CH$_3$OH, in low-mass protostars. For these sources, we  recalculate the CH$_3$OH column densities using the CH$_3$OH integrated line intensities from \citet{Oberg14a} and our rotation diagram method.

\section{Observations}

\begin{table*}[htp]
{
\begin{center}
\caption{Source information of the complete 16-object $c2d$ embedded protostar sample with ice detections} 
\label{Table1}
\begin{tabular}{l c c c c c c c c c c}
\hline\hline
Source & R.A. & Dec & Cloud & L$_{\rm bol}$	&M$_{\rm env}$	&$\alpha_{\rm IR}^{\rm a}$ &$N(H_2O)^{\rm a}$	&X$_{\rm CH_3OH}^{\rm b}$&X$_{\rm CH_4}^{\rm c}$\\
	&(J2000.0)	&(J2000.0)		&&L$_\odot$	&M$_\odot$	&&10$^{18}$ cm$^{-2}$	&\% H$_2$O&\% H$_2$O\\
\hline
B1-a$^{\rm d}$ 	&03:33:16.67	&31:07:55.1			&Perseus	&1.3$^{\rm e}$	&2.8$^{\rm e}$	&1.87	&10.39 [2.26]	&$<$1.9	&$<$5.7	\\
SVS 4-5$^{\rm d}$ 	&18:29:57.59	&01:13:00.6			&Serpens	&38$^{\rm f}$	& -- &1.26	&5.65 [1.13]	&25.2 [3.5]	&6.1 [1.7]	\\
B1-c		&03:33:17.89	&31:09:31.0	&Perseus	&3.7$^{\rm e}$& 17.7$^{\rm e}$&2.66	&29.55 [5.65]	&$<$7.1	&5.4 [1.4]	\\
IRAS 23238+7401	&23:25:46.65&74:17:37.2&CB244&--&--&0.95&12.95 [2.26] &	$<$3.6&	$<$7.4\\
L1455 IRS3	&03:28:00.41	&30:08:01.2	&Perseus	&0.32$^{\rm e}$& 0.2$^{\rm g}$&0.98	&0.92 [0.37]	&$<$12.5	&--	\\
B5 IRS 1$^{\rm d}$ 	&03:47:41.61	&32:51:43.8			&Perseus	&4.7$^{\rm e}$	&4.2$^{\rm e}$&0.78	&2.26 [0.28]	&$<$3.7	&--	\\
L1455 SMM1 &03:27:43.25	&30:12:28.8  &Perseus	&3.1$^{\rm e}$& 5.3$^{\rm e}$&2.41	&18.21 [2.82]	&$<$13.5	&5.8 [0.9]	\\
IRAS 03245+3002 & 03:27:39.03&30:12:59.3 &Perseus	&7.0$^{\rm e}$&5.3$^{\rm e}$&2.70	&39.31 [5.65]	&$<$9.8	&1.7 [0.3]	\\
L1014 IRS	&21:24:07.51&49:59:09.0&L1014&--&--&1.28&7.16 [0.91]&	3.1 [0.8]	&7.1	 [2.3]\\
IRAS 04108+2803$^{\rm d}$ 	&04:13:54.72	&28:11:32.9	&Taurus	&0.62$^{\rm h}$	&	-- &0.90	&2.87 [0.4]	&$<$3.5&	$<$11\\
IRAS 03235+3004$^{\rm d}$ 	&03:26:37.45	&30:15:27.9	&Perseus	&1.9$^{\rm e}$	&2.4$^{\rm e}$	&1.44	&14.48 [2.26] &	4.2 [1.2]	&4.3 [1.4]	\\
L1489 IRS$^{\rm d}$ &04:04:43.07	&26:18:56.4		&Taurus	&3.7$^{\rm h}$	& 0.1$^{\rm i}$&1.10	&4.26 [0.51]	&4.9 [1.5]	&3.1 [0.2]	\\
HH 300	&04:26:56.30	&24:43:35.3	&Taurus	&1.27$^{\rm h}$& 0.03$^{\rm j}$ &0.79	&2.59 [0.25]	&$<$6.7	&$<$14	\\
IRAS 03271+3013	&03:30:15.16	&30:23:48.8	&Perseus	&0.8$^{\rm e}$&1.2$^{\rm e}$&2.06	&7.69 [1.76] &$<$5.6	&$<$1.6	\\
L1448 IRS1 & 03:25:09.44 & 30:46:21.7 &Perseus	&17.0$^{\rm e}$ &16.3$^{\rm e}$ &0.34		&0.47 [0.16]	&$<$14.9	&--	\\ 
IRAS 03254+3050	&03:28:34.51	&31:00:51.2	&Perseus	&--	&0.3$^{\rm e}$	&0.90	&3.66 [0.47]	&$<$4.6	&4.0	\\
\hline
\hline
\end{tabular}
\end{center}
$^{\rm a}$\citet{Boogert08},
$^{\rm b}$\citet{Bottinelli10},
$^{\rm c}$\citet{Oberg08},
$^{\rm d}$Sources were observed by \citet{Oberg14a},
$^{\rm e}$\citet{Hatchell07},
$^{\rm f}$\citet{Pontoppidan04},
$^{\rm g}$\citet{Enoch09},
$^{\rm h}$\citet{Furlan08},
$^{\rm i}$\citet{Brinch07},
$^{\rm j}$\citet{Arce06}

}
\end{table*}

All sources were observed with the IRAM 30m telescope using the EMIR 90 GHz receiver and the Fourier Transform Spectrometer (FTS) backend. Six of the sources (B1-a, B5 IRS 1, L1489 IRS, IRAS 04108+2803, SVS 4-5, and IRAS 03235+3004)  were observed on June 12 -- 16, 2013 at 93 -- 101 GHz and 109 -- 117 GHz. The remaining sources were observed on July 23 -- 28, 2014 at 92 -- 100 GHz and 108 -- 116 GHz. The spectral resolution for both sets of observations was 200 kHz and the sideband rejection was $-15$ dB \citep{carter12}.

The pointing accuracy was checked every 1--2 hrs and found to be within 2$''$-3$''$. Focus, which was checked every 4 hrs, remained stable with corrections of $<0.4$ mm. For the data set obtained in June 2013, both the position switching and wobbler switching modes were used during observations, but the position switching spectra has been excluded from this paper because of severe baseline instabilities; only wobbler switching was used for the July 2014 observations with a wobbler throw of 2$'$. We exclude IRAS 03254+3050 from further analysis due to significant self-absorption in the molecular targets of this study. No other source displayed signs of self-absorption in CH$_3$OH or C$_4$H. We also inspected the shapes of stronger lines, e.g. CN, in individual integrations and found no self-absorption in any high-density tracing line. This indicates a lack of dense material in the off-position and therefore a low probability of CH$_3$OH and C$_4$H-emitting, material in the wobble off-positions. 

The spectra were reduced using CLASS\footnote{http://www.iram.fr/IRAMFR/GILDAS} with a global baseline fit to each 4 GHz spectral chunk using four to seven line-free windows. Individual scans were  baseline subtracted and averaged, and antenna temperature, $T_{a}^{*}$, was converted  to main beam temperature, $T_{mb}$, by applying forward and beam efficiency values of  0.95 and 0.81. Using literature source velocities, the spectra were converted to rest frequency with additional adjustments made based on the CH$_3$OH 2-1 ladder. The CH$_3$OH 2-1 ladder emission was compared with previous observations for a subset of sources and was found to agree within 10\% \citep{Oberg09a}.

\section{Results}

\begin{figure*}[htp]
\centering
\epsscale{1}
\plotone{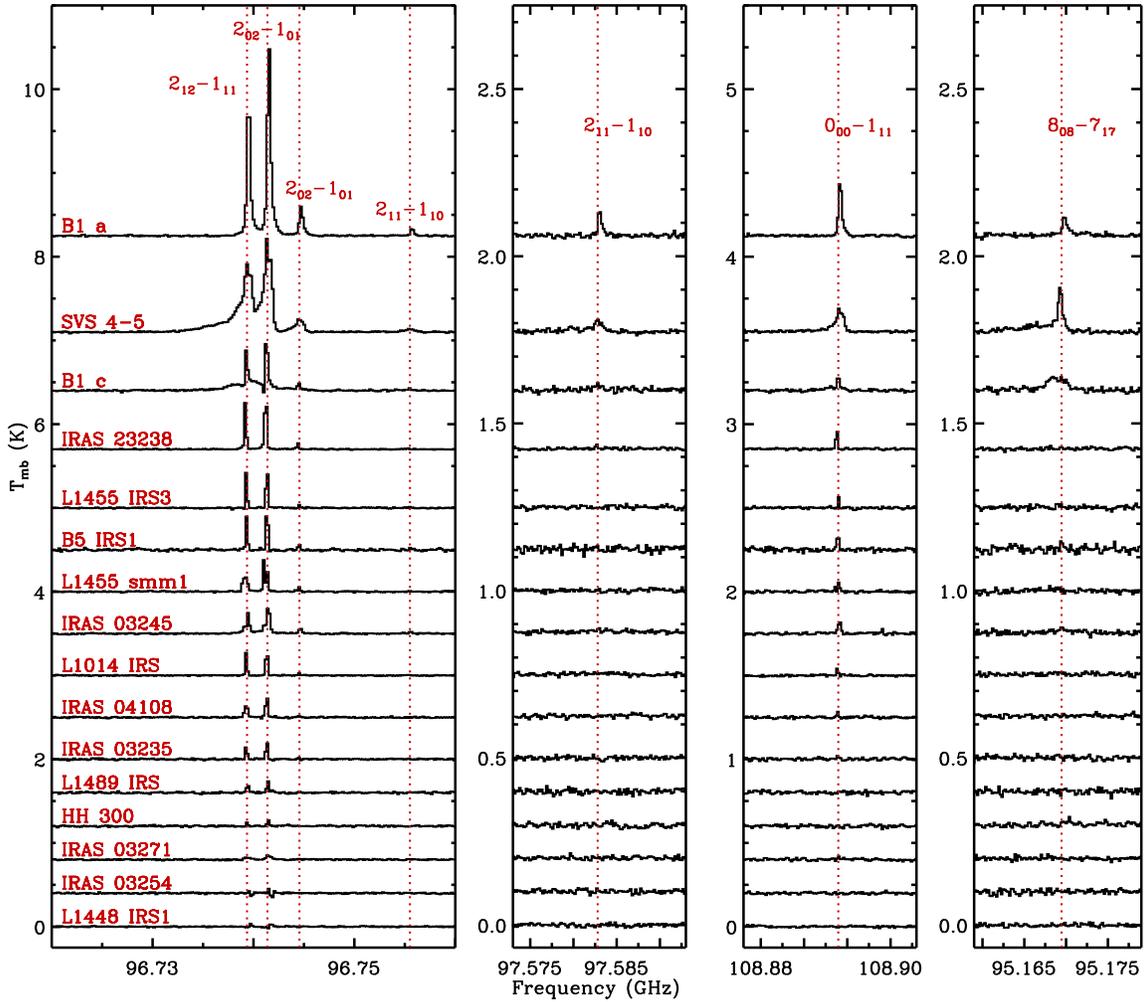}
\caption{IRAM 30m observations of the CH$_3$OH lines detected toward the low-mass YSO sample. The spectra have been shifted with the systemic velocity of each source.\label{CH3OH_spectra}}
\end{figure*}

\begin{figure*}[htp]
\centering
\epsscale{1}
\plotone{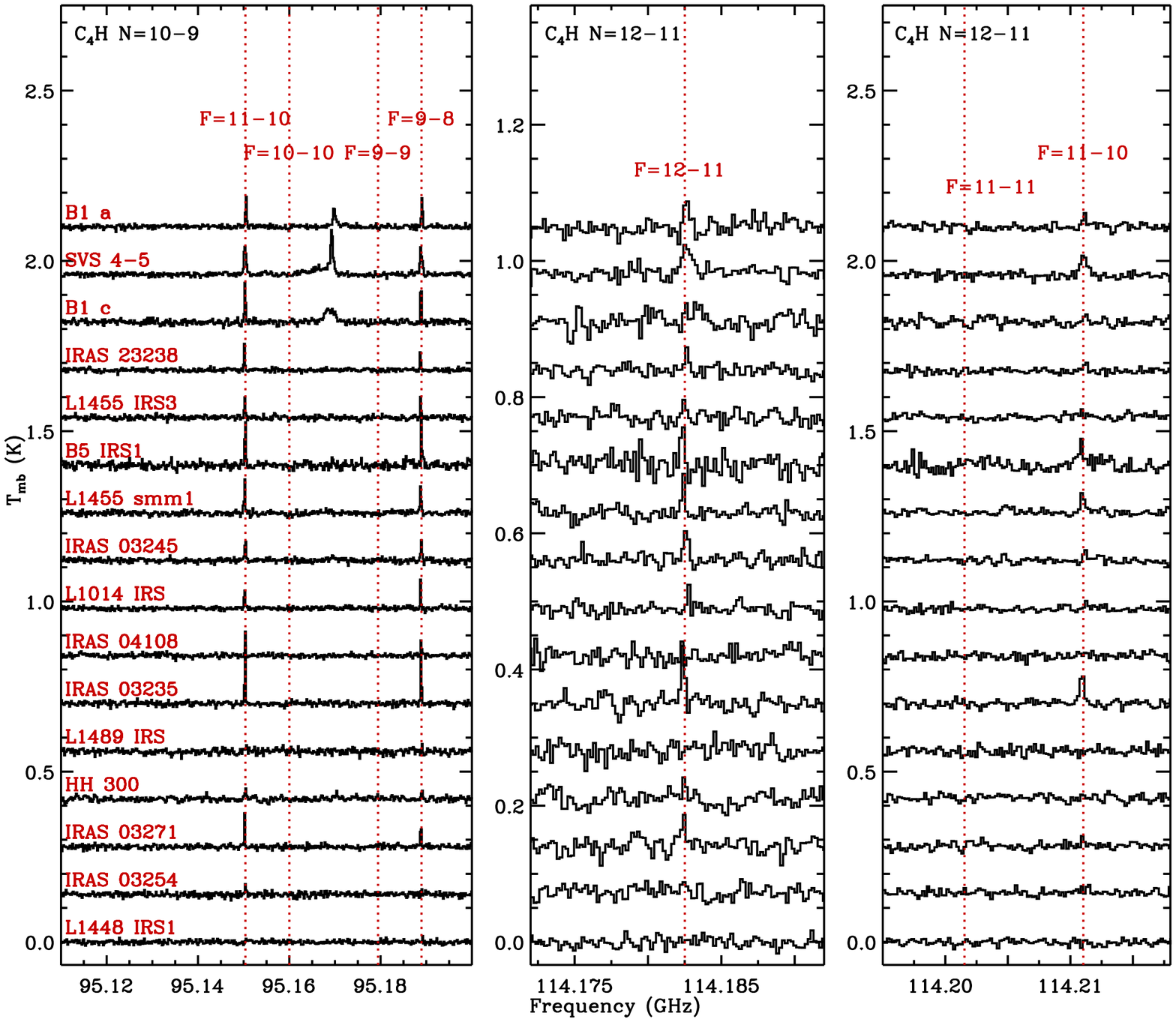}
\caption{IRAM 30m observations of the C$_4$H lines detected toward the low-mass YSO sample. The spectra have been shifted with the systemic velocity of each source.\label{C4H_spectra}}
\end{figure*}

\begin{figure*}[htp]
\centering
\epsscale{1}
\plotone{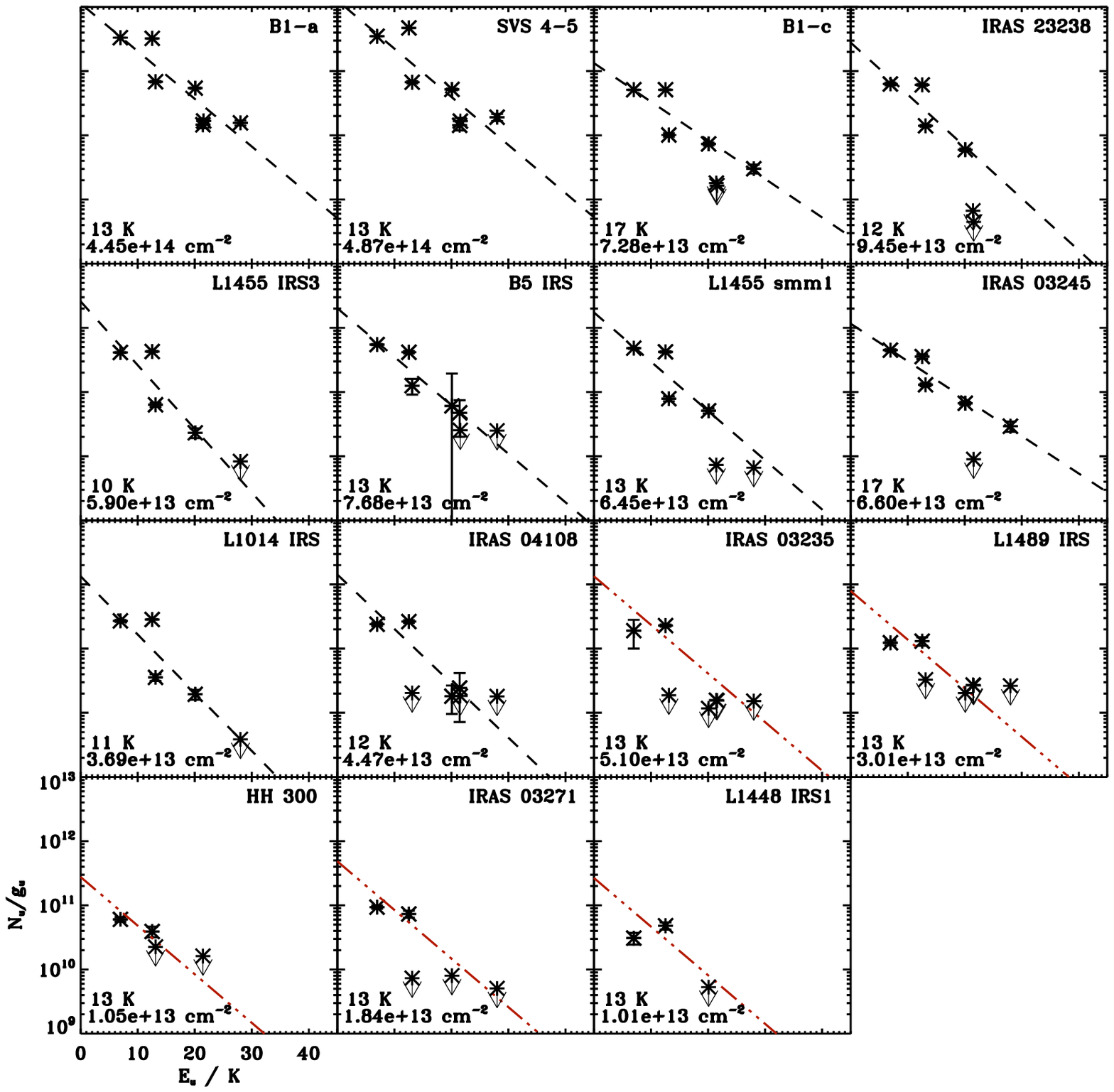}
\caption{Rotation Diagrams for CH$_3$OH. The black dashed lines represent the fit to the lines. When a line could not be fit, denoted by the red dot-dash line, the average rotational temperature was used to derive the column density. Where no error bars are seen, the errors bars are smaller than the symbols. \label{CH3OH rd}
}
\end{figure*}

\begin{figure*}[htp]
\centering
\epsscale{1}
\plotone{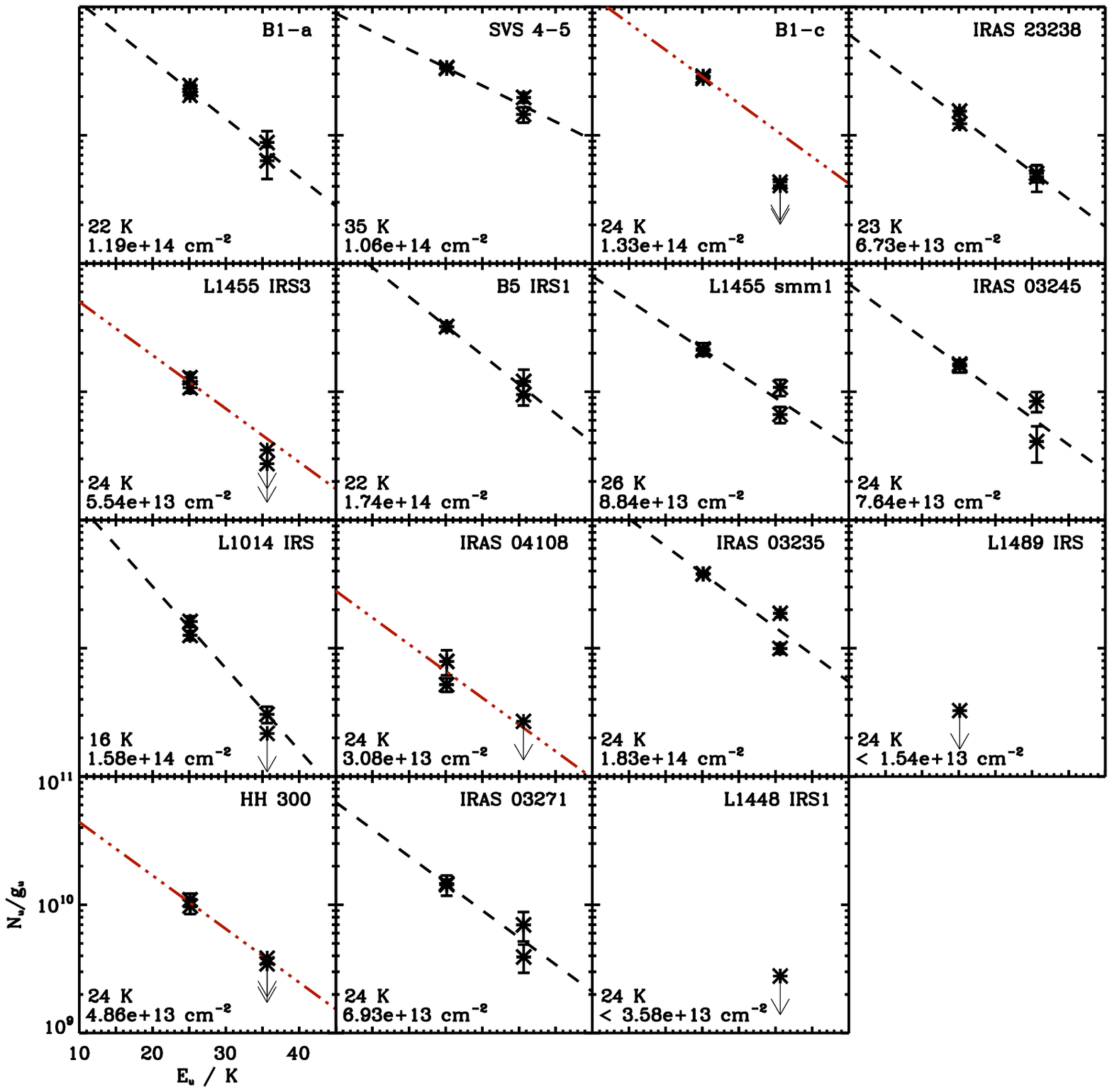}
\caption{Rotation Diagrams for C$_4$H. The black dashed lines represent the fit to the lines. When a line could not be fit, denoted by the red dot-dash line, the average rotational temperature was used to derive the column density. Where no error bars are seen, the errors bars are smaller than the symbols.\label{C4H rd}}
\end{figure*}

CH$_3$OH is detected in all 15 analyzed sources and C$_4$H in 13 sources at greater than 3$\sigma$. Two C$_4$H upper limits are also reported. Figures \ref{CH3OH_spectra} and \ref{C4H_spectra} display the CH$_3$OH and C$_4$H lines toward all observed sources. For ease of comparison, the CH$_3$OH lines in Figure \ref{CH3OH_spectra} are ordered in decreasing CH$_3$OH brightness from the top. Figure \ref{C4H_spectra} is ordered identically. Tables \ref{Methanol_integrated_lines}  and \ref{C4H_integrated} list the integrated intensity of the CH$_3$OH and C$_4$H lines, respectively. The integrated intensity of CH$_3$OH lines of sources studied by \citet{Oberg14a} can be found in Table 2 of that paper. Most CH$_3$OH and all C$_4$H lines could be fit by a single Gaussian. Toward SVS 4-5 and B1-c, the CH$_3$OH lines display substantial wings, probably due to outflows in the beam, and these features were fit with 2--3 Gaussian components. Only the narrow central Gaussian components were used to derived the listed column densities, since their properties agreed the best with the C$_4$H line characteristics in the same lines of sight.

\begin{table*}[htp]
{
\begin{center}
\caption{Integrated CH$_3$OH Line Intensities in K km s$^{-1}$} 
\label{Methanol_integrated_lines}
\begin{tabular}{l c c c c c c c}
\hline\hline
& 95.914 GHz& 96.739 GHz& 96.741 GHz& 96.745 GHz& 96.756 GHz& 97.583 GHz& 108.894 GHz\\
Source&2$_{12}$-1$_{11}$&2$_{12}$-1$_{11}$&2$_{02}$-1$_{01}$&2$_{02}$-1$_{01}$&2$_{11}$-1$_{10}$&2$_{11}$-1$_{10}$&0$_{00}$-1$_{11}$\\
\hline
B1-c & $<$0.013& 0.361[0.004] &0.481[0.004]& 0.069[0.004]& 0.022[0.003]& $<$0.012&0.065[0.004]\\
IRAS 23238+7401& $<$0.005& 0.430[0.011]&0.591[0.027]&0.056[0.002]&--&$<$0.003&0.090[0.003]\\
L1455 IRS3&--&0.300[0.007]&0.387[0.005]&0.022[0.004]&$<$0.006&--&0.040[0.003]\\
L1455 SMM1&$<$0.005&0.297[0.008]&0.452[0.009]&0.048[0.004]&$<$0.005&--&0.050[0.004]\\ 
IRAS 03245+3002& --& 0.250[0.021] & 0.421[0.015]& 0.063[0.003]& 0.021[0.004]& $<$0.006&0.083[0.004]\\
L1014 IRS& -- & 0.200[0.002]& 0.254[0.010]& 0.018[0.003]& $<$0.003& -- & 0.023[0.003]\\
HH 300& $<$0.011 & 0.027[0.005] & 0.057[0.005]& -- & --& --& $<$0.014\\
IRAS 03271+3013 & -- & 0.051[0.007]&0.088[0.005]&$<$0.008&$<$0.004&--&$<$0.005\\
L1448 IRS1& --& 0.034[0.004]&0.029[0.006]&$<$0.005&--&--&--\\
\hline
\hline
\end{tabular}
\end{center}}
\end{table*}

\begin{table*}[htp]
{
\begin{center}
\caption{Integrated C$_4$H Line Intensities in K km s$^{-1}$} 
\label{C4H_integrated}
\begin{tabular}{l  c c c c }
\hline\hline
& 95.150 GHz& 95.189 GHz& 114.183GHz & 114.221 GHz\\
Source&N=10-9, F=11-10& N=10-9, F=9-8&N=12-11, F=12-11&N=12-11, F=11-10\\
\hline
B1-a&0.088[0.005]&0.094[0.004]&0.054[0.012]&0.036[0.010]\\
SVS 4-5&0.142[0.007]&0.130[0.008]&0.090[0.013]&0.112[0.010]\\
B1-c& 0.119[0.008]&0.111[0.004]&$<$0.027&$<$0.023\\
IRAS 23238+7401&0.066[0.004]&0.047[0.003]&0.031[0.005]&0.027[0.006]\\
L1455 IRS3& 0.055[0.005]&0.041[0.004]&$<$0.017&$<$0.019\\
B5 IRS 1& 0.138[0.008]&0.124[0.010]&0.059[0.010]&0.068[0.016]\\
L1455 SMM1&0.092[0.010]&0.081[0.005]&0.041[0.006]&0.061[0.009]\\ 
IRAS 03245+3002&0.070[0.005]&0.061[0.007]&0.052[0.009]&0.023[0.007]\\ 
L1014 IRS& 0.054[0.005]&0.062[0.006]&0.19[0.003]&$<$0.012 \\
IRAS 04108+2803&0.022[0.003]&0.030[0.007]&$<$0.017&--\\
IRAS 03235+3004& 0.163[0.003]&0.147[0.004]&0.062[0.005]&0.106[0.007]\\
L1489 IRS&--&$<$0.013&--&--\\
HH 300&0.047[0.006]&0.038[0.005]&$<$0.021&$<$0.022\\
IRAS 03271+3013& 0.062[0.004]&0.055[0.010]&0.043[0.011]&0.022[0.006]\\ 
L1448 IRS1&--&--&--&$<$0.016\\
\hline
\hline
\end{tabular}
\end{center}}
\end{table*}

\begin{table*}[htp]
{
\begin{center}
\caption{Column Densities and Rotational Temperatures for CH$_3$OH and C$_4$H} 
\label{Table2}
\begin{tabular}{l c c c c c}
\hline\hline
Source & N(CH$_3$OH) & T$_{\rm rot}$(CH$_3$OH)$^{\rm a}$& &N(C$_4$H) & T$_{\rm rot}$(C$_4$H)$^{\rm a}$\\
        &10$^{13}$ cm$^{-2}$ & K & & 10$^{13}$ cm$^{-2}$ & K\\

\hline
B1-a$^{\rm b}$ & 44 [14] & 13 [3] &&11.9 [2.8]& 22 [4]\\
SVS 4-5$^{\rm b}$ & 49 [19] & 13 [4]&& 10.6 [2.1]& 35 [8]\\
B1-c        & 7.3 [2.2]& 17 [5]&& 13.3 [0.7]& {\it 24 [5]}\\
IRAS 23238+7401	& 9.4 [4.4]& 12 [5]&& 6.7 [1.0]& 23 [3]\\
L1455 IRS3 & 5.9 [3.6]& 10 [4]&& 5.5 [0.5]& {\it 24 [5]}\\
B5 IRS 1$^{\rm b}$ & 7.7 [2.0]& 13 [6]&& 17.4 [2.6]& 22 [2]\\
L1455 SMM1 & 6.4 [3.3]& 13 [6]&& 8.8 [2.8]& 26 [7]\\
IRAS 03245+3002 & 6.6 [1.2]& 17 [3]&& 7.6 [3.5]& 24 [9]\\
L1014 IRS & 3.7 [2.4]& 11 [6]&& 15.8 [4.2]& 16 [2]\\
IRAS 04108+2803$^{\rm b}$ & 4.5 [1.9]& 12 [4]&& 3.1 [0.4]& {\it 24 [5]}\\
IRAS 03235+3004$^{\rm b}$	& 5.1 [0.3]&{\it 13 [2]}&& 18.3 [7.4]& 24 [7]\\
L1489 IRS$^{\rm b}$ & 3.0 [0.6]& {\it 13 [2]}&& $<$1.5& {\it 24 [5]}\\
HH 300 & 1.0 [0.4]& {\it 13 [2]}&& 4.9[0.6]&{\it 24 [5]}\\
IRAS 03271+3013 & 1.8 [0.2]&{\it 13 [2]}&& 6.9 [2.5]& 27 [7]\\
L1448 IRS1 & 1.0 [0.2]&{\it 13 [2]}&& $<$3.6&{\it 24 [5]}\\
\hline
\hline
\end{tabular}
\end{center}
$^{\rm a}$T$_{\rm rot}$ values in italics are assumed rotational temperatures and are based on the average T$_{\rm rot}$ in the source with standard deviation errors. The listed errors are the 1$\sigma$ uncertainty.
$^{\rm b}$Sources observed by \citet{Oberg14a}
}
\end{table*}

To determine the column densities of CH$_3$OH and C$_4$H, we used the rotation diagram method  of \citet{Goldsmith99} and assumed optically thin lines and local thermodynamic equilibrium (LTE) at a single temperature. Figures \ref{CH3OH rd} and \ref{C4H rd} display the rotation diagrams obtained for CH$_3$OH and C$_4$H, respectively.  Sources observed by \citet{Oberg14a} were refit using a single-component rotation diagram to be consistent with the column density determinations of C$_4$H and the new CH$_3$OH observations.

Table \ref{Table2} lists the derived column densities and rotational temperature obtained from the rotation diagram analysis. In situations where there were not enough lines with SNR $>$ 3$\sigma$, the rotational temperature is assumed to be the average of the observed rotational temperatures; in these cases, T$_{\rm rot}$ is italicized.   On average, T$_{\rm rot}$  for CH$_3$OH is 13 $\pm$ 2 K and 24  $\pm$ 5 K for C$_4$H.  It is important to note that for this class of objects the CH$_3$OH rotational temperatures do not necessarily reflect the kinetic temperatures, but are rather lower limits \citep{Oberg14a}. The difference in excitation temperature for CH$_3$OH and C$_4$H can thus not be used to constrain the relative emission regions of CH$_3$OH and C$_4$H.

\begin{figure}[htp]
\centering
\epsscale{0.8}
\plotone{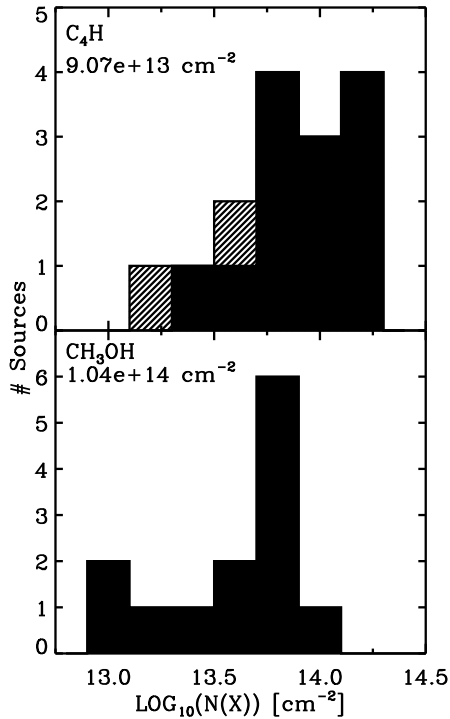}
\caption{Histograms of the column density of  C$_4$H and CH$_3$OH. The solid bars are detections and line-filled bars indicate upper limits. The average column density is also listed in the upper left hand corner.\label{histo}}
\end{figure}

The observed column density distributions of C$_4$H and CH$_3$OH are displayed in Figure \ref{histo} along with their respective average column densities. The column densities of both species span approximately an order of magnitude and the average column densities of 9.1 $\times$ 10$^{13}$ cm$^{-2}$ and 1.0 $\times$ 10$^{14}$ cm$^{-2}$  for C$_4$H and CH$_3$OH, respectively are similar. The WCCC sources from \citet{Sakai08, Sakai09b, Sakai09a}  overlap with the high end of our C$_4$H distribution. Classical hot corino sources \citep{Maret05, Sakai09b} overlap with the high-end of the CH$_3$OH distribution and the low end of our distribution for C$_4$H. However, as is seen below, classical hot corino sources form a distinct sub-group when simultaneously taking into account the column densities of both species, compared to any sources in our sample.

\begin{figure*}[htp]
\centering
\epsscale{0.9}
\plotone{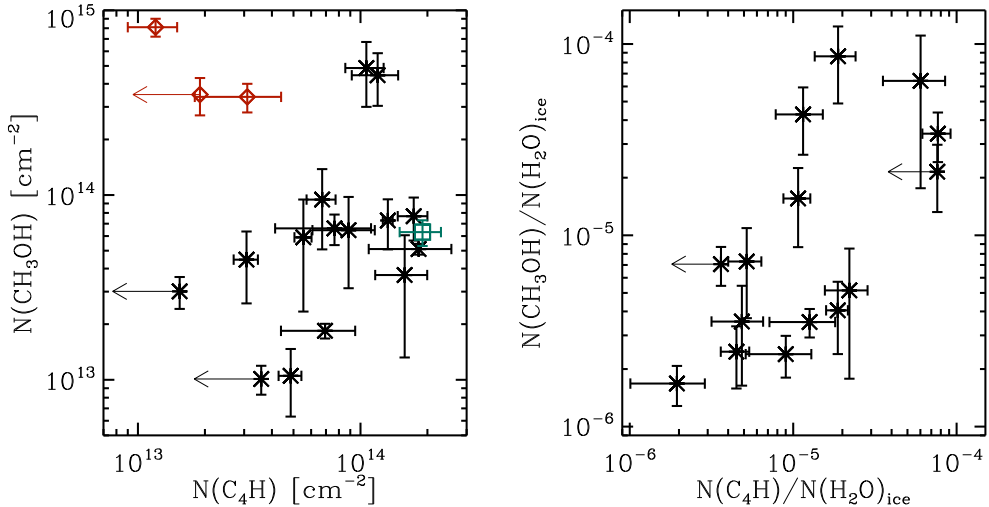}
\caption{Correlation plots for the column density of C$_4$H and CH$_3$OH from our sample. The left plot displays the pure column density correlation where the black data points are our data, the teal square is L1527 \citep{Sakai08, Sakai09a}, and the red diamonds are three hot corino sources \citep{Maret05, Sakai09a}. The right plot displays the CH$_3$OH to C$_4$H correlation following a normalization using the  H$_2$O ice column density.\label{C4H_v_CH3OH}}
\end{figure*}

Figure \ref{C4H_v_CH3OH} displays the column density correlation plots for CH$_3$OH and C$_4$H, both absolute and normalized to the H$_2$O ice column. As H$_2$O ice is the first ice to form and the last to desorb during star-formation, its column should be a good proxy of the envelope material in the line of sight \citep{Boogert15}. 

For both column density correlation plots, there appears to be a positive correlation. To determine the statistical significance of the correlation, a Spearman's (rho) rank correlation tests were performed. The absolute column densities are not significantly correlated. By contrast, the normalized (to water ice) column densities are significantly positively correlated at the $>$99\% level applying the Spearman's rank correlation test.

In Figure \ref{C4H_v_CH3OH},  the warm carbon chain source, L1527, is displayed as a teal square and falls within our observed columns for both CH$_3$OH and C$_4$H \citep{Sakai08,Sakai09a}. The red diamonds display three classic hot corino sources \citep{Maret05, Sakai09a}. The three hot corino sources and the WCCC source L1527 in Figure \ref{C4H_v_CH3OH} display an anti-correlation. This is consistent with the anti-correlation noted by \citet{Sakai08}. When combing the hot corino and WCCC sources with our sample, we see two distinct regions:  a positive correlation for low to medium column densities of CH$_3$OH, while the hot corino sources occupy a unique space characterized by high CH$_3$OH and low C$_4$H column densities.

The difference between our sample and the hot corino sources suggest that CH$_3$OH/C$_4$H may trace an evolutionary sequence. We used the IR spectral indices, $\alpha_{\rm IR}$, as a proxy for age since $\alpha_{\rm IR}$ decreasing signifies an increasing age \citep{Wilking01}. We find no relationship for the CH$_3$OH/C$_4$H ratio to $\alpha_{\rm IR}$, indicating that there is no evolutionary sequence within our sample. Instead observed variations in the CH$_3$OH/C$_4$H ratio in our sample (visible in Figure \ref{C4H_v_CH3OH} as a large spread in CH$_3$OH column densities for any specific C$_4$H column density) seems to be the result of different initial ice compositions.
 
\begin{figure}[htp]
\centering
\epsscale{0.9}
\plotone{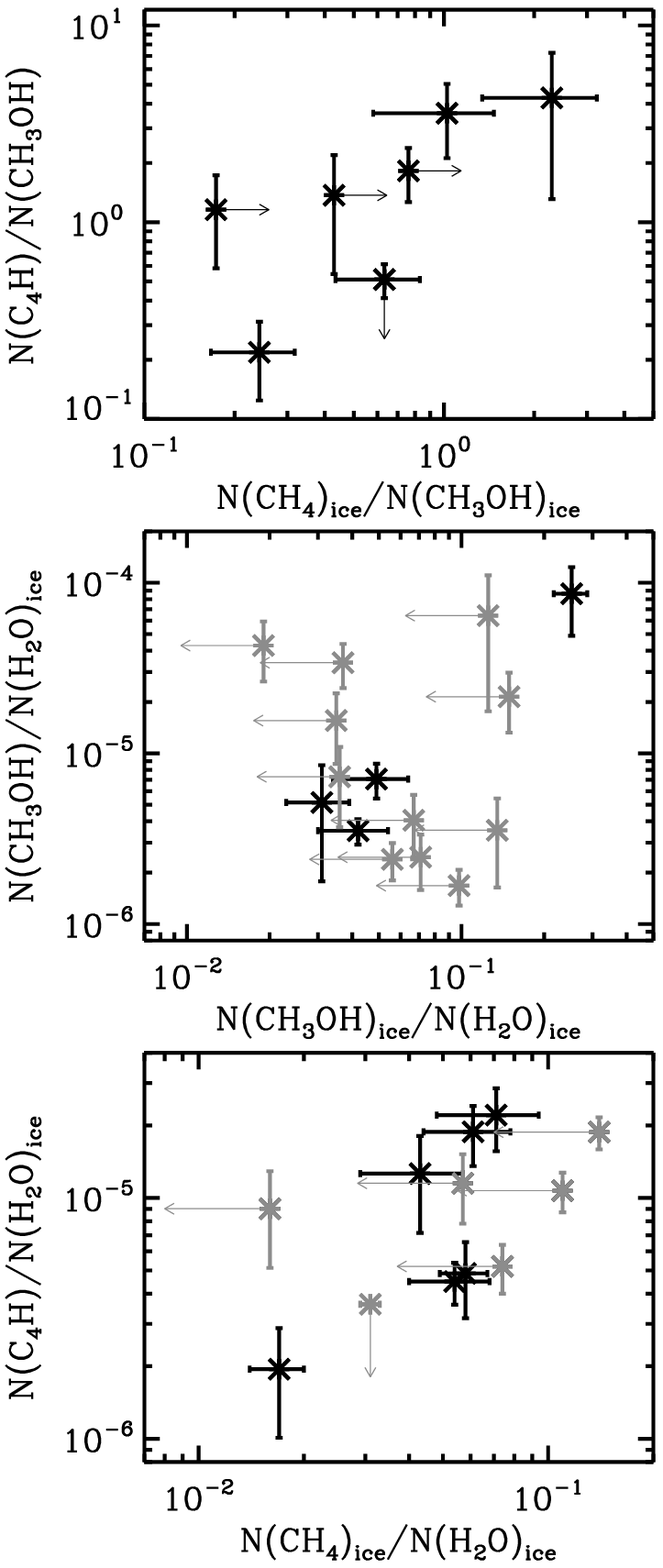}
\caption{Ice and gas correlation plots from our sample. The top plot displays the CH$_4$ to CH$_3$OH ice ratio versus the C$_4$H to CH$_3$OH gas ratio. The middle plot displays the ice CH$_3$OH versus the gas CH$_3$OH  columns, and the bottom plot displays the ice CH$_4$ versus the gas C$_4$H columns. Both of the bottom two plots are normalized to the H$_2$O ice column density and the gray data points indicate the upper limits. \label{ices}}
\end{figure}

Figure \ref{ices} shows a positive correlation between the C$_4$H/CH$_3$OH  gas and the ice phase CH$_4$/CH$_3$OH. The correlation is not statistically significant due to the many ice upper limits, however it does suggest a relationship between gas and ice abundances for these species. When C$_4$H gas and CH$_4$ ice, and CH$_3$OH ice and gas are considered separately, there seems to be a stronger relationship between the gas and ice phase hydrocarbons (Figure \ref{ices} bottom) than between CH$_3$OH gas and ice (Figure\ref{ices} middle).

\section{Discussion}

In this study, we observed 16 embedded protostars to determine the relationship between CH$_3$OH and C$_4$H. We find that there is a positive correlation between the column density of CH$_3$OH and C$_4$H. Several of the sources have COM detections in addition to CH$_3$OH \citep{Oberg14a}.
WCCC and COM chemistry are thus not mutually exclusive during the embedded stages of star formation. Rather, carbon chain and COM formation seem to follow one another up until the onset of an efficient hot corino chemistry. Once the hot corino has formed,  CH$_3$OH and C$_4$H are no longer correlated \citep{Maret05}.

\citet{Aikawa12} modeled the chemistry from a prestellar core up to the formation of the protostellar envelope, and found that both carbon chains and COMs can be present in the colder, outer envelope. \citet{Aikawa12} attribute the formation of the carbon chains to an increase in the C$^+$ abundance in conjunction with the CH$_4$ ice sublimation. The presence of CH$_3$OH in the envelope is due to non-thermal sublimation of CH$_3$OH ice. Contrary to C$_4$H, CH$_3$OH is also abundant toward the protostellar core due to thermal sublimation. The abundance of CH$_3$OH in the protostellar core can be many orders of magnitude higher than the outer envelope emission, but its contribution to observed emission can be quite negligible if the volume of thermally sublimated CH$_3$OH is small compared to the protostellar envelope volume. Should the emission from the thermally sublimated CH$_3$OH dominate the CH$_3$OH envelope emission, the source would be considered a hot corino.

Within our source sample, envelope ice abundances explain variations in the C$_4$H/CH$_3$OH ratio. The CH$_3$OH ice and gas columns (Figure \ref{ices} middle) are not clearly related, however. If all CH$_3$OH in all sources originated from non-thermal desorption there should be a correlation between CH$_3$OH gas and ice abundances. On the other hand if some CH$_3$OH is due to thermal desorption, the source luminosity will influence the CH$_3$OH emission, muddling the expected CH$_3$OH gas and ice correlation. As we observe no correlation, it is likely that we are probing both thermal and non-thermal desorption processes.

The observed tentative C$_4$H-CH$_{\rm 4,ice}$ correlation indicates that the formation of C$_4$H depends on the initial abundance of CH$_4$ ice. The thermal sublimation of CH$_4$ ice to promote WCCC was initially suggested by \citet{Sakai08}, and is consistent with current chemical models \citep{Hassel11, Aikawa12}. Not all C$_4$H may originate from this process, however, as some may be inherited from the molecular cloud. An unusually high contribution of inherited C$_4$H,  may explain the outlier with a high C$_4$H gas column and low CH$_4$ ice upper limit (IRAS 03271+3010) in Figure \ref{ices}. A possible explanation for such a high contribution may be that this source is very young.

In addition to ice abundances, the chemistry of C$_4$H and CH$_3$OH may depend on environmental parameters, such as envelope mass and bolometric luminosity. A higher bolometric luminosity should result in a larger hot corino region. If this is the main regulator for the COMs-carbon chain relationship, the CH$_3$OH/C$_4$H ratio would increase toward more luminous protostars. We do not find any correlation between this ratio and source luminosity (Tables \ref{Table1} and \ref{Table2}). It is more difficult to assess the importance of the initial envelope mass on the chemistry, since  envelope mass evolves with time. A current low envelope mass measurement may signify either a low-mass protostar or an older more massive protostar. We do not find any correlation between the CH$_3$OH/C$_4$H ratio and envelope mass alone (Tables \ref{Table1} and \ref{Table2}). The small sample size prevented a more detailed analysis of the combined influence of mass and age on the chemistry. We also do not see any relationships between these environmental parameters and the absolute column densities of C$_4$H and CH$_3$OH.

As reported in \S4, we also find no correlation with $\alpha_{\rm IR}$, an age indicator, and the CH$_3$OH/C$_4$H  ratio. For young embedded protostars, the `age' of the system does not seem to affect the relative importance of carbon chain and COM chemistry, i.e. C$_4$H and CH$_3$OH coexist in the envelope at this evolutionary stage.  In Figure \ref{C4H_v_CH3OH}, the ``boomerang" shape may indicate an evolutionary trend when comparing these young objects with more evolved ones. This is consistent with models. As the protostar transitions into the hot corino stage, the temperature is high enough to both destroy carbon chains and thermally desorb CH$_3$OH, resulting in an anti-correlation in the observed C$_4$H and CH$_3$OH column densities \citep{Aikawa12}. 

\section{Conclusions}

We surveyed 16 northern sources selected from the Spitzer $c2d$ ice sample using the IRAM 30m and found the following:

\begin{enumerate}
\item C$_4$H and CH$_3$OH coexist and are correlated at the deeply embedded stage of low-mass protostellar evolution. This can be explained by lukewarm environments in the protostellar envelopes that promote WCCC alongside non-thermal desorption of CH$_3$OH. 

\item The CH$_3$OH/C$_4$H correlation does not extend into the hot corino phase. This is indicative of an evolutionary sequence where carbon chains and COMs coexist in lukewarm protostellar envelopes, but once a hot corino forms, COMs are enhanced and carbon chains are destroyed.

\item The ice and gas abundances are related in these sources based on a tentative correlation between CH$_3$OH/C$_4$H gas ratio and CH$_3$OH/CH$_4$ ice ratio. It is likely that the C$_4$H gas and CH$_4$ ice correlation is what is driving this relationship, suggesting that WCCC is intimately connected to the CH$_4$ ice abundance.

\end{enumerate}

\acknowledgements  This work has benefited from discussions with Ryan Loomis, Viviana Guzman, Edith Fayolle, and from the helpful comments of the anonymous referee. The study is  based on observations with the IRAM 30m Telescope. IRAM is supported by INSU/CNRS (France), MPG (Germany) and IGN (Spain). KI\"O acknowledges funding from the Simons Collaboration on the Origins of Life Investigator award \#321183, the Alfred P. Sloan Foundation, and the David and Lucile Packard Foundation.


\end{document}